\documentclass[pre,twocolumn,aps,showpacs]{revtex4}
\newcommand{\bec}[1]{\mbox{\boldmath $ #1$}}
\newcommand{\Ra}{\mathcal{R}a}
\usepackage{verbatim}
\usepackage{graphicx}
\begin{document}
\title{Large-scale instabilities in a non-rotating turbulent convection}
\author{Tov Elperin}
\email{elperin@bgu.ac.il} \homepage{http://www.bgu.ac.il/~elperin}
\author{Ilia Golubev}
\email{golubev@bgu.ac.il}
\author{Nathan Kleeorin}
\email{nat@bgu.ac.il}
\author{Igor Rogachevskii}
\email{gary@bgu.ac.il}
\homepage{http://www.bgu.ac.il/~gary}
\affiliation{Pearlstone Center for Aeronautical Engineering Studies,
Department of Mechanical Engineering, Ben-Gurion University of the
Negev, POB 653, Beer-Sheva 84105, Israel}
\date{\today}
\begin{abstract}
A theoretical approach proposed in Phys. Rev. E {\bf 66}, 066305
(2002) is developed further to investigate formation of large-scale
coherent structures in a non-rotating turbulent convection via
excitation of a large-scale instability. In particular, the
convective-wind instability that causes formation of large-scale
coherent motions in the form of cells, can be excited in a
shear-free regime. It was shown that the redistribution of the
turbulent heat flux due to non-uniform large-scale motions plays a
crucial role in the formation of the coherent large-scale structures
in the turbulent convection. The modification of the turbulent heat
flux results in strong reduction of the critical Rayleigh number
(based on the eddy viscosity and turbulent temperature diffusivity)
required for the excitation of the convective-wind instability. The
large-scale convective-shear instability that results in the
formation of the large-scale coherent motions in the form of rolls
stretched along imposed large-scale velocity, can be excited in the
sheared turbulent convection. This instability causes the generation
of convective-shear waves propagating perpendicular to the
convective rolls. The mean-field equations which describe the
convective-wind and convective-shear instabilities, are solved
numerically. We determine the key parameters that affect formation
of the large-scale coherent structures in the turbulent convection.
In particular, the degree of thermal anisotropy and the lateral
background heat flux strongly modify the growth rates of the
large-scale convective-shear instability, the frequencies of the
generated convective-shear waves and change the thresholds required
for the excitation of the large-scale instabilities. This study
elucidates the origins of the large-scale circulations and rolls in
the atmospheric convective boundary layers and the meso-granular
structures in the solar convection.
\end{abstract}


\maketitle

\section{Introduction}

Large-scale coherent structures in a non-rotating turbulent
convection at very large Rayleigh numbers are observed in the
atmospheric convective boundary layers
\cite{EB93,AZ96,LS80,H84,W87,H88,SS89,Z91,WK96,ZGH98,BR99,YKH02}, in
numerous laboratory experiments in the Rayleigh-B\'{e}nard apparatus
\cite{KH81,SWL89,CCL96,K01,NSS01,NS03,BKT03,XL04,SQ04,BNA05,EEKR06}
and in direct numerical simulations \cite{HTB03,PHP04}. Spatial
scales of the large-scale coherent structures in a turbulent
convection are much larger than turbulent scales and their life-time
is larger than the largest time scales of turbulence. In the
atmospheric shear-free convection, the structures (cloud cells)
represent large, three-dimensional, long-lived B\'{e}nard-type cells
composed of narrow uprising plumes and wide downdraughts. They
usually embrace the entire convective boundary layer (of the order
of 1-3 km in height) and include pronounced convergence flow
patterns close to the surface. In the sheared convective flows, the
structures represent large-scale rolls (cloud streets) stretched
along the mean wind \cite{EB93,AZ96,YKH02}.

Coherent structures in convective turbulent flows were
comprehensively studied theoretically, experimentally and in
numerical simulations
\cite{EB93,AZ96,LS80,H84,W87,H88,SS89,Z91,WK96,ZGH98,BR99,
YKH02,KH81,SWL89,CCL96,K01,NSS01,NS03,BKT03,XL04,SQ04,BNA05,EEKR06,HTB03,PHP04,BW71,
BW74,B83}. However, some aspects related to the origin of
large-scale coherent structures in non-rotating turbulent convection
are not completely understood. Hartlep et al. (2003) noted that
there are two points of view on the origin of large-scale
circulation in turbulent convection \cite{HTB03}. According to one
point of view, the rolls which develop at low Rayleigh numbers near
the onset of convection continually increase their size as Rayleigh
number is increased and continue to exist in an average sense at
even the highest Rayleigh numbers reached in the experiments
\cite{F76}. Another hypothesis holds that the large-scale
circulation is a genuine high Rayleigh number effect \cite{KH81}.

Recently, a new mean-field theory of non-rotating turbulent
convection has been developed \cite{EKRZ02,EKRZ06}. This theory
predicts the convective-wind instability in the shear-free turbulent
convection that results in the formation of large-scale motions in
the form of cells. In the sheared convection, the large-scale
instability causes generation of convective-shear waves. The
dominant coherent structures in this case are rolls. It was
demonstrated \cite{EKRZ02,EKRZ06} that a redistribution of the
turbulent heat flux due to non-uniform large-scale motions plays a
crucial role in the formation of the large-scale coherent structures
in turbulent convection.

In this study a theoretical approach \cite{EKRZ02,EKRZ06} is
developed further to investigate the formation of the coherent
structures in the non-rotating turbulent convection. In particular,
we investigated how the modification of the turbulent heat flux due
to non-uniform large-scale motions affects the critical Rayleigh
number (based on the eddy viscosity and turbulent thermal
conductivity) required for the excitation of the convective-wind
instability. We performed numerical study of the convective-wind and
convective-shear instabilities in order to determine key parameters
that affect formation of the large-scale coherent structures in the
turbulent convection.

The paper is organized as follows. In Section~II we discussed the
physics of the formation of the large-scale coherent structures and
formulated the mean-field equations which describe the formation of
the coherent structures. In Section~III we determined the critical
Rayleigh number required for the excitation of the convective-wind
instability in a shear-free turbulent convection. In Section~IV we
studied numerically the convective-shear instability in a sheared
convection. Finally, conclusions are drawn in Section~V.

\section{Turbulent heat flux and mean-field equations}

In this Section we discuss a redistribution of the turbulent heat
flux due to the non-uniform large-scale motions as a key mechanism
for the formation of the large-scale coherent structures  in
turbulent convection. Here we also formulate the mean-field
equations which describe the formation of the coherent structures.
Traditional theoretical models of the boundary-layer turbulence,
such as the Kolmogorov-type local closures, imply the following
assumptions. Fluid flows are decomposed into organized mean motions
and turbulent flow. Turbulent fluxes are assumed to be proportional
to the local mean gradients, whereas the proportionality
coefficients (eddy viscosity, turbulent temperature diffusivity) are
uniquely determined by local turbulent parameters. For example
\cite{MY75}, the turbulent heat flux reads ${\bf F} \equiv \langle s
\, {\bf u} \rangle = - \kappa_{_{T}} \bec{\nabla}  S$, where
$\kappa_{_{T}}$ is the turbulent temperature diffusivity, $S$ is the
mean entropy, ${\bf u}$ and $s$ are fluctuations of the velocity and
entropy, respectively. This turbulent heat flux ${\bf F}$ does not
include the contribution from anisotropic velocity fluctuations.

Actually the mean velocity gradients can directly affect the
turbulent heat flux. The reason is that additional essentially
anisotropic velocity fluctuations are generated by tangling the
mean-velocity gradients with the Kolmogorov-type turbulence due to
the influence of the inertial forces during the life time of large
turbulent eddies. The Kolmogorov-type turbulence supplies energy to
the anisotropic velocity fluctuations which cause formation of
coherent structures due to the excitation of a large-scale
instability \cite{EKRZ02,EKRZ06}. Anisotropic velocity fluctuations
are characterized by a steeper spectrum than the Kolmogorov-type
turbulence \cite{L67,WC72,SV94,IY02,EKRZ02}.

The theoretical model \cite{EKRZ02} of the anisotropic velocity
fluctuations and their effect on the turbulent heat flux includes
the following steps in the derivations: applying the spectral
closure, solving the equations for the second moments in the ${\bf
k}$ space, and returning to the physical space to obtain formulas
for the Reynolds stresses and the turbulent heat flux. The
derivation are based on the Navier-Stokes equation and the entropy
evolution equation formulated in the Boussinesq approximation. This
derivation \cite{EKRZ02} yields the following expression for the
turbulent heat flux ${\bf F} \equiv \langle s \, {\bf u} \rangle$:
\begin{eqnarray}
{\bf F} &=& {\bf F}^{\ast} -  \tau_0 \, \biggl[\alpha \, {\bf
F}_{z}^{\ast} \, {\rm div} \, {\bf U}_{\perp} - {1 \over 5} \,
\biggl(\alpha + {3 \over 2} \biggr) \, ( {\bf W} {\bf \times} {\bf
F}_z^{\ast})
\nonumber \\
& & - {1 \over 2} ( {\bf W}_{z} {\bf \times} {\bf F}^{\ast}) \biggr]
\;, \label{A1}
\end{eqnarray}
where $\tau_0$ is the correlation time of turbulent velocity
corresponding to the maximum scale of turbulent motions, $ {\bf W} =
\bec{\nabla} {\bf \times}  {\bf U}$ is the mean vorticity, $ {\bf U}
=  {\bf U}_{\perp} +  {\bf U}_{z}$ is the mean velocity with the
horizontal $ {\bf U}_{\perp}$ and vertical $ {\bf U}_{z}$
components, $\alpha$ is the degree of thermal anisotropy, and
\begin{eqnarray}
F_i^{\ast} = - \kappa_{ij} {\nabla}_j  S - \tau_0 \, F_z^{\ast} \,
{\nabla}_z \,  U_i^{(0)}(z) \label{A2}
\end{eqnarray}
is the background turbulent heat flux that is the sum of the
contribution due to the Kolmogorov-type turbulence (described by the
first term in Eq.~(\ref{A2})) and a contribution of the anisotropic
turbulence caused by the shear of the imposed large-scale mean
velocity ${\bf U}^{(0)}(z)$ (the so-called counter-wind heat flux
described by the second term in Eq.~(\ref{A2})),
\begin{eqnarray}
\kappa_{ij} = \kappa_{_{T}} \, [\delta_{ij} + b\, e_i \, e_j]
\label{A3}
\end{eqnarray}
is a generalized anisotropic turbulent temperature diffusivity
tensor. For turbulent convection  $b = (3/2) \, (2+\tilde \gamma)$,
$\, \tilde \gamma$ is the ratio of specific heats (e.g., $\tilde
\gamma = 7/5$ for the air flow) and ${\bf e}$ is the vertical unit
vector. The equation for the tensor $\kappa_{ij}$ was derived in
Appendix A in Ref. \cite{EKRZ02} using the budget equations for the
turbulent kinetic energy, fluctuations of the entropy and the
turbulent heat flux. The anisotropic part of the tensor
$\kappa_{ij}$ (described by the second term in the square brackets
of Eq.~(\ref{A3})), is caused by a modification of the turbulent
heat flux by the buoyancy effects. Note that for a laminar
convection $b$ is set to be zero and the coefficient of the
turbulent temperature diffusivity $\kappa_{_{T}}$ should be replaced
by the coefficient of the molecular temperature diffusivity. The
parameter $\alpha$ in Eq.~(\ref{A1}) is given by
\begin{eqnarray}
\alpha &=& {1 + 4 \xi \over 1 + \xi / 3} \;, \quad  \quad \xi  =
\biggl({l_{\perp} \over l_{z}} \biggr)^{2/3} - 1 \;,
\label{A4}
\end{eqnarray}
where $l_{\perp}$ and $l_{z}$ are the horizontal and vertical scales
in which the background turbulent heat flux $F_{z}^{\ast}({\bf r}) =
\langle s({\bf x}) \, u_z({\bf x}+{\bf r}) \rangle$ tends to zero.
The parameter $\xi$ describes the degree of thermal anisotropy. In
particular, in isotropic case when $l_{\perp} = l_{z}$ the parameter
$\xi = 0$ and $\alpha = 1$. For $l_{\perp} \ll l_{z}$ the parameter
$\xi = - 1$ and $\alpha = - 9/2$. The maximum value $ \xi_{\rm max}
$ of the parameter $\xi$ is given by $\xi_{\rm max} = 2/3$ for
$\alpha = 3$. The upper limit for the parameter $\xi$ arises because
the function $F_{z}^{\ast}({\bf r})$ has a global maximum at ${\bf
r}=0$. Depending on the parameter $\alpha$ the small-scale thermal
structures in the background turbulent convection have the form of
columns or pancakes (sometimes they are called as the small-scale
thermal plumes). For $\alpha < 1$ the small-scale thermal structures
have the form of columns $(l_{\perp} < l_{z})$, and for $\alpha > 1$
there exist the pancake thermal structures $(l_{\perp}
> l_{z})$ in the background turbulent convection (i.e., a turbulent
convection with zero gradients of the mean velocity).

The terms in the square brackets in the right hand side of
Eq.~(\ref{A1}) result from the anisotropic turbulence and depend on
the "mean" (including coherent) velocity gradients. These terms lead
to the excitation of large-scale instability and formation of
coherent structures. In Eq.~(\ref{A1}) the terms with zero
divergence are omitted, because only ${\rm div} \, {\bf F}$
contributes to the mean-field dynamics. Neglecting the anisotropic
turbulence term in Eq.~(\ref{A1}) recovers the traditional equation
for the turbulent heat flux.

The physical meaning of Eq.~(\ref{A1}) is the following. The first
term $\propto - \tau_0 \, \alpha \, {\bf F}_{z}^{\ast} \, {\rm div}
\, \tilde {\bf U}_{\perp}$ in square brackets in Eq.~(\ref{A1})
describes the redistribution of the vertical background turbulent
heat flux ${\bf F}_{z}^{\ast}$ by the perturbations of the
convergent (or divergent) horizontal mean flows $\tilde {\bf
U}_{\perp}$. This redistribution of the vertical turbulent heat flux
occurs during the life-time of turbulent eddies. The second term
$\propto \tau_0 \, (\alpha + 3 /2) \, (\tilde {\bf W} {\bf \times}
{\bf F}_z^{\ast})$ in square brackets in Eq.~(\ref{A1}) determines
the formation of the horizontal turbulent heat flux due to
"rotation" of the vertical background turbulent heat flux ${\bf
F}_{z}^{\ast}$  by the perturbations of the horizontal mean
vorticity $\tilde {\bf W}_{\perp}$. The last term $\propto \tau_0 \,
(\tilde {\bf W}_{z} {\bf \times} {\bf F}^{\ast})$ in square brackets
in Eq.~(\ref{A1}) describes the formation of the horizontal heat
flux through the "rotation" of the horizontal background heat flux
${\bf F}_{\perp}^{\ast}$ (the "counter-wind" heat flux in
Eq.~(\ref{A2})) by the perturbations of the vertical component of
the mean vorticity $\tilde {\bf W}_{z}$. These three effects are
determined by the local inertial forces in inhomogeneous mean flows.
A more detailed discussion of Eq.~(\ref{A1}) is given in
Sections~III-IV.

The counter-wind turbulent heat flux (in the direction opposite to
the mean wind) is well-known in the atmospheric physics and arises
due to the following reason. In the sheared turbulent convection an
ascending fluid element has larger temperature then that of
surrounding fluid and smaller horizontal fluid velocity, while a
descending fluid element has smaller temperature and larger
horizontal fluid velocity. This causes the background turbulent heat
flux ${\bf F}_{\perp}^{\ast} = - \tau_0 \, F_z^{\ast} \, {\nabla}_z
\, {\bf U}_{\perp}^{(0)}(z)$ in the direction opposite to the
background horizontal mean sheared fluid velocity ${\bf
U}_{\perp}^{(0)}(z)$.

We use the mean-field approach whereby the small-scale turbulent
convection is parameterized. This is a reason why we do not use
explicitly thermal plumes in the consideration. The main reason for
the appearance of the large-scale coherent structures is related to
the modification of the heat flux by the non-uniform mean flows. The
thermal plumes contribute to the modification of the turbulent heat
flux. To a some extent, the redistribution of the turbulent heat
flux can be interpreted as a redistribution of the thermal plumes.

In order to study the formation of the large-scale coherent
structures in a small-scale non-rotating turbulent convection we
used the mean-field Navier-Stokes equation and the mean entropy
evolution equation (with the turbulent heat flux~(\ref{A1}))
formulated in the Boussinesq approximation. These mean-field
equations yield the following linearized equations for the small
perturbations from the equilibrium, $ \tilde U = U_{z} - U_{z}^{(0)}
,$ $ \, \tilde W = W_z -  W_z^{(0)} $ and $ \tilde S = S -  S^{(0)}
$:
\begin{eqnarray}
&& \biggl({\partial  \over \partial t} +  U_{y}^{(0)} \, \nabla_y -
\nu_{_{T}} \, \Delta \biggr) \, \Delta \, \tilde U = g \,
\Delta_{\perp} \, \tilde S \;,
\nonumber\\
\label{B3} \\
&& \biggl({\partial  \over \partial t} +  U_{y}^{(0)} \, \nabla_y -
\nu_{_{T}} \, \Delta \biggr) \, \tilde W = - \sigma \, \nabla_x \,
\tilde U \;,
\label{B4} \\
&& \biggl({\partial  \over \partial t} +  U_{y}^{(0)} \, \nabla_y
\biggr) \, \tilde S = - (\bec{\nabla} \cdot \tilde {\bf F}) -
(\nabla_z S^{(0)}) \, \tilde U \;,
 \label{B5}
\end{eqnarray}
where $\nu_{_{T}}$ is the eddy viscosity, $\Delta_{\perp} = \Delta -
\nabla^2_z ,$ and
\begin{eqnarray}
\bec{\nabla} \cdot \tilde {\bf F} = - {4 \tau_{0} \over 45} \biggl[
({\bf F}^{\ast} \cdot {\bf e}) [10 \, \alpha \, \Delta_{\perp} - (8
\, \alpha - 3) \Delta] \, \tilde U
\nonumber\\
+ 6 \, [({\bf F}^{\ast} {\bf \times} {\bf e}) \cdot \bec{\nabla}] \,
\tilde W \biggr] - \kappa_{_{T}} \, (\Delta + b \, \nabla^2_z) \,
\tilde S \;, \label{B1}
\end{eqnarray}
$\kappa_{_{T}}$ is the turbulent temperature diffusivity. In order
to derive Eq.~(\ref{B3}), the pressure term was excluded by
calculating the curl of the momentum equation. Equations
(\ref{B3})-(\ref{B5}) allows us to study the linear stage of the
large-scale instabilities. The variables $\tilde U$, $\, \tilde W$
and $\tilde S$ describe the large-scale coherent structures. In
Section~III we study a shear-free convection with ${\bf U}^{(0)} =
0$, and in Section~IV we investigate turbulent convection with the
background (equilibrium) large-scale velocity shear ${\bf
U}^{(0)}(z) = \sigma \, z \, {\bf e}_{y}$ and the background mean
vorticity ${\bf W}^{(0)} = \bec{\nabla} {\bf \times} {\bf U}^{(0)} =
- \sigma \, {\bf e}_x$.

\section{Shear-free convection}

Let us consider a shear-free convection $( {\bf U}^{(0)} = 0)$. In
the shear-free regime, the large-scale instability is related to the
first term $\propto - \tau_0 \, \alpha \, {\bf F}_{z}^{\ast} \, {\rm
div} \, \tilde{\bf U}_{\perp}$ in square brackets in Eq.~(\ref{A1})
for the turbulent heat flux \cite{EKRZ02,EKRZ06}. When $\partial
\tilde U_{z} /\partial z > 0$, perturbations of the vertical
velocity $\tilde U_{z}$ cause negative divergence of the horizontal
velocity, ${\rm div} \, \tilde {\bf U}_{\perp} < 0 $ (provided that
${\rm div} \, \tilde {\bf U} = 0). $ This strengthens the local
vertical turbulent heat flux and causes increase of perturbations of
the local mean entropy and buoyancy. The latter enhances
perturbations of the local mean vertical velocity $\tilde U_{z}$,
and by this means the convective-wind instability is excited.
Similar reasoning is valid when $\partial \tilde U_{z} /\partial z <
0$, whereas ${\rm div} \, \tilde {\bf U}_{\perp} > 0 $. Then
negative perturbations of the vertical flux of entropy leads to a
decrease of perturbations of the mean entropy and buoyancy, that
enhances the downward flow and once again excites the
convective-wind instability. Therefore, nonzero ${\rm div} \, \tilde
{\bf U}_{\perp}$ causes redistribution of the vertical turbulent
heat flux and formation of regions with large values of this flux.
These regions (where ${\rm div} \, \tilde {\bf U}_{\perp} < 0 $)
alternate with the low heat flux regions (where ${\rm div} \, \tilde
{\bf U}_{\perp} > 0 $). This process results in formation of the
large-scale coherent structures.

The role of the second term $\propto \tau_0 \, (\alpha + 3 /2) \,
(\tilde {\bf W} {\bf \times} {\bf F}_z^{\ast})$ in square brackets
in Eq.~(\ref{A1}) is to decrease the growth rate of the large-scale
instability for $\alpha > - 3/2$. Indeed, the interaction of
perturbations of the mean vorticity with the vertical background
turbulent heat flux ${\bf F}_z^{\ast}$ produces the horizontal
turbulent heat flux. The latter decreases (increases) the mean
entropy in the regions with upward (downward) local flows, thus
diminishing the buoyancy forces and reducing the mean vertical
velocity $\tilde U_{z}$ and the mean vorticity $\tilde {\bf W}$.
This mechanism dampens the convective-wind instability for $\alpha >
- 3/2$. The above two competitive effects determine the growth rate
of the convective-wind instability. A solution of Eqs.~(\ref{B3})
and (\ref{B5}) in the shear-free convection regime yields the
following expression for the growth rate $\gamma$ of long-wave
perturbations:
\begin{eqnarray}
\gamma \propto g \, F_z^\ast \, \tau_0^2 \, K^{2} \, \sqrt{\beta} \,
| \sin \theta | \, \biggl[\alpha - {3 \over 8} - {5 \alpha \over 4}
\,  \sin^{2} \theta \biggr]^{1/2} \;, \label{A5}
\end{eqnarray}
where the parameter $\beta=(l_0 \, K)^{-2} \gg 1$, $\, l_0$  is the
maximum scale of turbulent motions, $\theta$ is the angle between
the vertical unit vector ${\bf e}$ and the wave vector ${\bf K}$ of
small perturbations, ${\bf g}$ is acceleration of gravity. The
analysis of the convective-wind instability was performed in Refs.
\cite{EKRZ02,EKRZ06} only for a small square of
Brunt-V\"{a}is\"{a}l\"{a} frequency. In particular, Eq.~(\ref{A5})
was derived in Refs. \cite{EKRZ02,EKRZ06} for the case $|N^2| \ll g
F_z^\ast \tau_0 K^{2}$, where $N^2 = - ({\bf g} \cdot \bec{\nabla})
\, S^{(0)} < 0$ is the square of Brunt-V\"{a}is\"{a}l\"{a}
frequency.

In the present study we consider arbitrary values of the
Brunt-V\"{a}is\"{a}l\"{a} frequency, and we investigate the effect
of the modification of the turbulent heat flux (due to non-uniform
large-scale motions) on the critical effective Rayleigh number
required for the excitation of the convective-wind instability. We
also study here the effect of the anisotropy of turbulent thermal
diffusivity (caused by the buoyancy) on the critical effective
Rayleigh number. To this end we rewrite Eqs.~(\ref{B3})
and~(\ref{B5}) in a non-dimensional form
\begin{eqnarray}
\bigg({\partial \over \partial t} - \Delta\bigg) \, \Delta \, V &=&
{\Ra} \, \Delta_\bot \, \tilde S \;,
\label{C5}\\
{\rm Pr}_{_{\rm T}} \,  {\partial \tilde S \over \partial t} -
(\Delta + b \, \nabla^2_z) \, \tilde S  &=& V + \mu \, \biggl({{\rm
Pr}_{_{\rm T}} \over {\Ra}}\biggr)^{1/3} \, \Big[10 \, \alpha \,
\Delta_\bot
\nonumber \\
&& - (8 \, \alpha - 3) \, \Delta \Big] \, V \;,
\label{C6}
\end{eqnarray}
where $V = {\rm Pr}_{_{\rm T}} \, \zeta \, \Lambda \, \tilde U $ is
dimensionless velocity, the length is measured in the units of the
total vertical size $L_z$ of the system, the parameter $b=3 \,
(2+\tilde \gamma) / 2$ describes the anisotropy of turbulent thermal
diffusivity caused by the buoyancy effect (see Eqs.~(\ref{A3})), $\,
{\Ra} = 6 \, \Lambda^3 \, \zeta \, {\rm Pr}_{_{\rm T}}$ is effective
Rayleigh number based on the turbulent viscosity, $\nu_{_{T}}$, and
the turbulent temperature diffusivity, $\kappa_{_{T}}$, $\, \Lambda=
L_z /l_0$, $\, {\rm Pr}_{_{\rm T}} = \nu_{_{T}} / \kappa_{_{T}}$ is
the turbulent Prandtl number, the parameter $\mu$ is given by
\begin{eqnarray*}
\mu=\frac{4 \,  a_{\ast}}{15 } \, \biggl({6 \over
\zeta^{2}}\biggr)^{1/3} \;, \quad \zeta = \frac{6 \, g \,
l_0}{u_0^2} \, \frac{\delta T}{T_0} \;,
\end{eqnarray*}
$\delta T$ is the mean temperature difference between bottom and
upper boundaries of the turbulent convection, the parameter
$a_{\ast}=2 \, g \, \tau_0 \, F_z^{\ast} / u_0^2$ and $T_0$ is the
reference mean temperature. The last term $\propto \mu$ in the right
hand side of Eq.~(\ref{C6}) determines the modification of the
turbulent heat flux due to the non-uniform large-scale motions, and
the parameter
\begin{eqnarray*}
\mu \, \biggl({ {\rm Pr}_{_{\rm T}} \over {\Ra}}\biggr)^{1/3} = {4
\, g \, \tau_0 \, F_z^{\ast} \over |N^2| L_z^2}
\end{eqnarray*}
has the meaning of the normalized heat flux, where $|N^2| = g \,
\delta T / [T_0 \, L_z]$ and ${\Ra} = |N^2| L_z^4 / [ \nu_{_{T}} \,
\kappa_{_{T}}]$.

\subsection{Solution for two free boundaries}

\begin{figure}
\centering
\includegraphics[width=7cm]{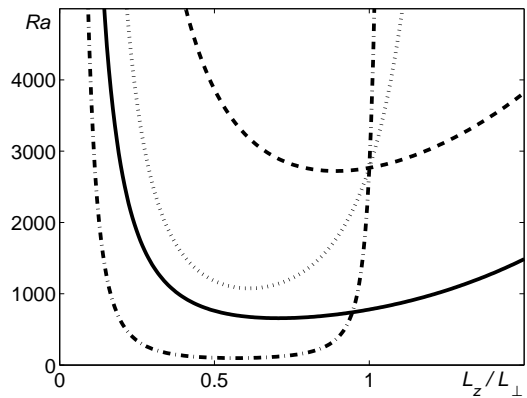}
\caption{\label{FIG1} Effective Rayleigh number versus the aspect
ratio $L_z/L_{\perp}$ of the perturbations for two free boundaries
and different values of the parameter $\mu$: $\, \, \, \mu=0$
(dashed line); for $\mu = 0.7$ (dotted line); $\mu = 5$
(dashed-dotted line). Here $\alpha=1$ and $b = 5.1$. The classical
Rayleigh solution for the laminar convection $(b = 0)$ with the two
free boundaries is shown by solid curve.}
\end{figure}

Let us consider the solution of Eqs.~(\ref{C5}) and (\ref{C6}) for
two free boundaries, using the following boundary conditions
\begin{eqnarray}
V &=& \nabla^2_z \, V = \tilde S = 0 \qquad \textrm{for} \qquad z =
0; \, 1 \; . \label{M1}
\end{eqnarray}
We seek for a solution of Eqs.~(\ref{C5})-(\ref{C6})  in the form
\begin{eqnarray*}
V, \, \tilde S \propto \sin (\pi \, n \, z) \, \exp(\gamma \, t - i
\, {\bf K}_{\perp} \cdot {\bf r}) \;,
\end{eqnarray*}
where $n$ is the integer number and $K_{\perp}$ is the horizontal
component of the wave vector. The critical effective Rayleigh number
(at $\gamma=0$) is determined by the equation
\begin{eqnarray*}
&&(K^2_{\perp} + (b+1)\, \pi^2 \, n^2) \, (K^2_{\perp} + \pi^2 \,
n^2)^2 =  K^2_{\perp} \, \Big[{\Ra}_c
\nonumber \\
&&-  \mu \, ({\rm Pr}_{_{\rm T}} \, {\Ra}_c^{2})^{1/3} \, \big[(2
\alpha+3) \, K^2_{\perp} -(8 \alpha-3) \, \pi^2 \, n^2 \big]
\Big]\,,
\end{eqnarray*}
where the critical effective Rayleigh number, ${\Ra}_c$, is based on
the turbulent viscosity and the turbulent temperature diffusivity.

In the case of $\mu=0$ (i.e., there is no modification of the
turbulent heat flux due to the non-uniform large-scale motions), the
critical effective Rayleigh number is given by
\begin{eqnarray}
{\Ra}_c = \frac{(K^2_{\perp} + \pi^2 \, n^2)^2 (K^2_{\perp} + (b+1)
\, \pi^2 \, n^2)}{K^2_{\perp}}.
\end{eqnarray}
The minimum value of the critical effective Rayleigh number for the
first mode $(n = 1)$ for $b=0$ is ${\Ra}_c \approx 657.5$. This is
the classical Rayleigh solution for the laminar convection with two
free boundaries. The critical effective Rayleigh number increases
with the increase of the anisotropy of turbulent temperature
diffusivity (see Eqs.~(\ref{A3})) described by the parameter $b$.
Indeed, for $b=3.9$ the critical effective Rayleigh number is
${\Ra}_c \approx 2247$ and for $b=5.1$ it is ${\Ra}_c \approx 2722$.

The modification of the turbulent heat flux due to non-uniform
large-scale motions strongly decreases the critical effective
Rayleigh number. Indeed, Fig.~\ref{FIG1} shows the effective
Rayleigh number versus the aspect ratio $L_{z} / L_{\perp} \equiv
K_{\perp} / K_{z} = \tan \theta$ of the perturbations for different
values of parameter $\mu$. The increase of the parameter $\mu$
causes strong reduction of the critical Rayleigh number (see
Table~\ref{Tab1}).

\subsection{Solution for two rigid boundaries}

\begin{figure}
\centering
\includegraphics[width=7cm]{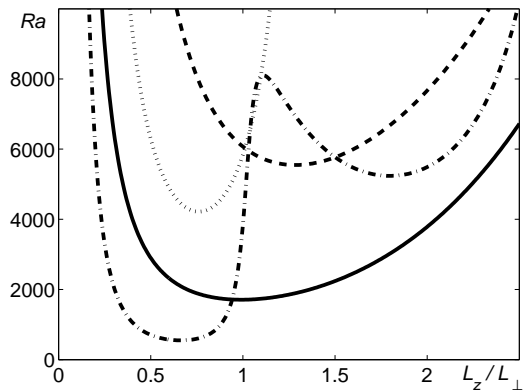}
\caption{\label{FIG2} Effective Rayleigh number versus the aspect
ratio $L_z/L_{\perp}$  of the perturbations for two rigid boundaries
and different values of the parameter $\mu$: $\, \, \, \mu = 0$
(dashed line); $\mu = 0.7$ (dotted line); $\mu = 5$ (dashed-dotted
line). Here $\alpha=1$ and $b = 5.1$. The classical Rayleigh
solution for the laminar convection $(b = 0)$ with the two rigid
boundaries is shown by solid curve.}
\end{figure}

\begin{figure}
\centering
\includegraphics[width=7cm]{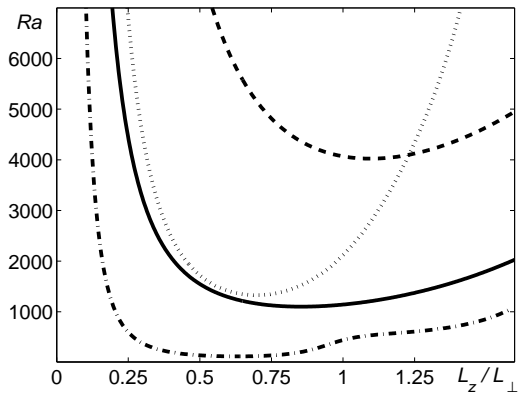}
\caption{\label{FIG3} Effective Rayleigh number versus the aspect
ratio $L_z/L_{\perp}$  of the perturbations for one rigid and one
free boundaries  and different values of the parameter $\mu$: $\, \,
\, \mu = 0$ (dashed line); $\mu = 0.7$ (dotted line); $\mu = 5$
(dashed-dotted line). Here $\alpha=1$ and $b = 5.1$. The classical
Rayleigh solution for the laminar convection $(b = 0)$ with the one
rigid and one free boundaries is shown by solid curve.}
\end{figure}

Now let us consider the solution of Eqs.~(\ref{C5}) and (\ref{C6})
for two rigid boundaries. In view of the symmetry of this problem
with respect to two bounding planes it is convenient to translate
the origin of $z$ to be midway between the two planes. Then fluid is
confined between two planes $z=\pm 1/2$, and we seek for a solution
of Eqs.~(\ref{C5}) and (\ref{C6}) satisfying the following boundary
conditions
\begin{eqnarray}
V &=& \nabla_z \, V = \tilde S = 0 \qquad \textrm{for} \qquad z =
\pm {1 \over 2} \; . \label{C2}
\end{eqnarray}
We seek for the solution of Eqs.~(\ref{C5}) and (\ref{C6}) in the
form $V, \, \tilde S~\propto~\exp(\gamma \, t  + q \, z - i \, {\bf
K}_{\perp} \cdot {\bf r})$, where the critical effective Rayleigh
number is determined by the equation
\begin{eqnarray}
&& [K_{\perp}^2 - (1 + b) \, q^2] \, (K_{\perp}^2 - q^2)^2  =
K_{\perp}^2 \, \Big[{\Ra}_c
\nonumber \\
&& - \mu \, ({\rm Pr}_{_{\rm T}} \, {\Ra}_c^{2})^{1/3} \, \big[(2
\alpha + 3) \, K_{\perp}^2 + (8 \alpha - 3) \,  q^2\big] \Big]
 \; .
 \nonumber \\
 \label{C3}
\end{eqnarray}
The problem is symmetric with respect to the two boundaries so the
eigenfunctions fall into two distinct classes: the even mode with
vertical velocity symmetry with respect to the mid plane and the odd
mode with vertical velocity asymmetry. Following the procedure
described in \cite{RH58,CH61,DR02}, we adopted the even solution
which has minimum critical effective Rayleigh number. Our numerical
analysis showed that the anisotropy of turbulent temperature
diffusivity described by the parameter $b$ increases the critical
effective Rayleigh number. In particular, for $b=0$ the critical
effective Rayleigh number is ${\Ra}_c \approx 1707.8$ This is
classical Rayleigh solution for the laminar convection with the two
rigid boundaries. For $\mu=0$ and $b = 3.9$ the critical effective
Rayleigh number is ${\Ra}_c \approx 4683$, and for $b = 5.1$ it is
${\Ra}_c \approx 5547$.

The effective Rayleigh number versus the aspect ratio
$L_z/L_{\perp}$  of the perturbations for two rigid boundaries is
plotted in Fig.~\ref{FIG2}. Increasing of the parameter $\mu$
decreases both, the critical effective Rayleigh number and the
aspect ratio $L_z/L_{\perp}$ of perturbations (see
Table~\ref{Tab1}). If $\mu \geq 5$, the behavior of the effective
Rayleigh number drastically changes, e.g., there are two local
minima for the effective Rayleigh number.

\subsection{Solution for one rigid and one free boundaries}

Solution for one rigid and one free boundaries can be obtained from
the results for two rigid boundaries using the odd mode. We use the
domain from $z=0$ (the free boundary) to $z=1/2$ (the rigid
boundary). The anisotropy of turbulent temperature diffusivity
described by the parameter $b$ increases the critical effective
Rayleigh number. Indeed, for $b=0$ the critical effective Rayleigh
number is ${\Ra}_c \approx 1101$ (the classical Rayleigh solution
for the laminar convection). For $\mu=0$ and $b = 3.9$ the critical
effective Rayleigh number is ${\Ra}_c \approx 3359$, and for $b =
5.1$ it is ${\Ra}_c \approx 4023$. The effective Rayleigh number
versus the aspect ratio $L_z/L_{\perp}$ of the perturbations for the
one rigid and one free boundaries is plotted in Fig.~\ref{FIG3}.
Increasing of the parameter $\mu$ decreases the critical effective
Rayleigh number and reduces the aspect ratio $L_z/L_{\perp}$  of the
perturbations (see Table~\ref{Tab1}).

Therefore, for these three types of boundaries the modification of
the turbulent heat flux due to the non-uniform large-scale motions
strongly reduces the critical effective Rayleigh number (based on
the eddy viscosity and turbulent temperature diffusivity) required
for the excitation of the convective-wind instability. We summarized
the final results for the above three types of the boundary
conditions in Table~\ref{Tab1}. The case of laminar convection is
presented in Table~\ref{Tab1} only for comparison with the results
obtained for the turbulent convection.

\begin{widetext}
\begin{table}
  \begin{tabular}{c c c c c c c}
  \hline \hline
 & \multicolumn{6}{c}{Boundaries} \\\cline{2-7}
 Case & \multicolumn{2}{c}{\qquad two free \qquad \qquad}
 &\multicolumn{2}{c}{one free and one rigid}&
 \multicolumn{2}{c}{\qquad two rigid \qquad \qquad}
    \\\cline{2-7}
    & $L_z/L_{\perp}$ & ${\Ra}_c$ & $L_z/L_{\perp}$ &
    ${\Ra}_c$& $L_z/L_{\perp}$ & ${\Ra}_c$ \\
    \hline
    Laminar flow: & 0.707 & 657.5 & 0.854 & 1101 & 0.994 & 1708 \\
    Turbulent flow: &       &       &      &       &       \\
     $\mu = 0$\quad & 0.891 & 2722 & 1.096 & 4023 & 1.280 & 5547 \\
    $\mu=0.7$ & 0.613 & 1076 & 0.697 & 1328 & 0.754 & 4218 \\
    $\mu=2.0$ & 0.578 & 344 & 0.645 & 420 & 0.688 & 1743 \\
    $\mu=5.0$ & 0.568 & 98 & 0.628 & 120 & 0.662 & 549 \\
    \hline \hline
  \end{tabular}
  \caption{Critical effective Rayleigh numbers for different types of
  the boundaries. Here for the turbulent flow $\alpha=1$ and $b = 5.1$.
  The case of laminar convection is presented in Table only for
  comparison with the results obtained for the turbulent convection.}
  \label{Tab1}
 \end{table}
\end{widetext}
\noindent

\bigskip

\section{Sheared turbulent convection}

In this Section we consider turbulent convection with a large-scale
linear velocity shear ${\bf U}^{(0)}(z) = \sigma \, z \, {\bf
e}_{y}$. In a sheared turbulent convection the mechanism of the
convective-shear instability \cite{EKRZ02,EKRZ06} is related to the
last term $\propto \tau_0 \, (\tilde {\bf W}_{z} {\bf \times} {\bf
F}^{\ast})$ in square brackets in Eq.~(\ref{A1}). The generation of
the potential temperature perturbations by vorticity perturbations
plays the key role in this mechanism. Indeed, in two adjacent
vortices with the opposite directions of the vertical vorticity
$\tilde {\bf W}_z$, the turbulent fluxes of potential temperature
are directed towards the boundary between the vortices. This
increases perturbations of the mean potential temperature and the
buoyancy, and generates the upward flow between the vortices. These
vertical flows excite vorticity perturbations, and the
convective-shear instability mechanism is sustained.

\begin{figure}
\vspace*{2mm}
\centering
\includegraphics[width=7cm]{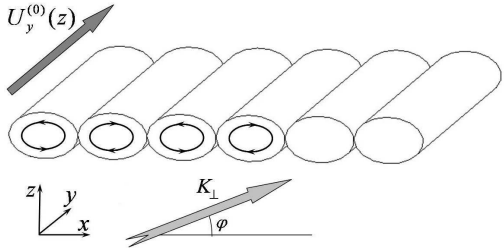}
\caption{\label{FIG4} Large-scale coherent rolls formed due to the
convective-shear instability and aligned along the sheared mean
velocity $ {\bf U}^{(0)}(z) .$ The instability results in generation
of the convective-shear waves which propagate perpendicular to the
convective rolls.}
\end{figure}

\begin{figure}
\centering
\includegraphics[width=7cm] {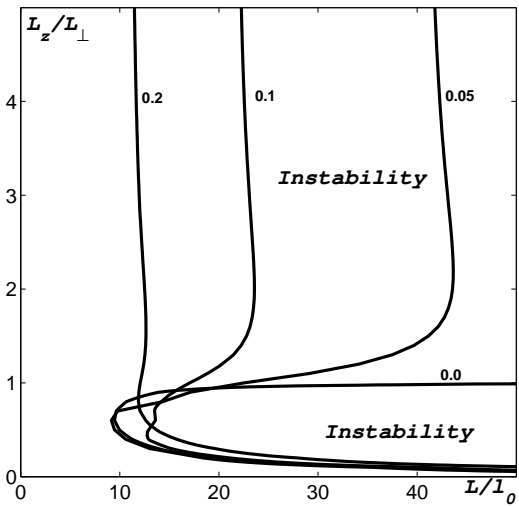}
\caption{\label{FIG5}Range of parameters ($L_z / L_{\perp}$; $L /
l_0$) for which the convective-shear instability occurs, for
different values of the shear parameter $\lambda$=0; 0.05; 0.1;
0.2\,. Here $\alpha=1$\,.}
\end{figure}

\begin{figure}
\centering
\includegraphics[width=7cm]{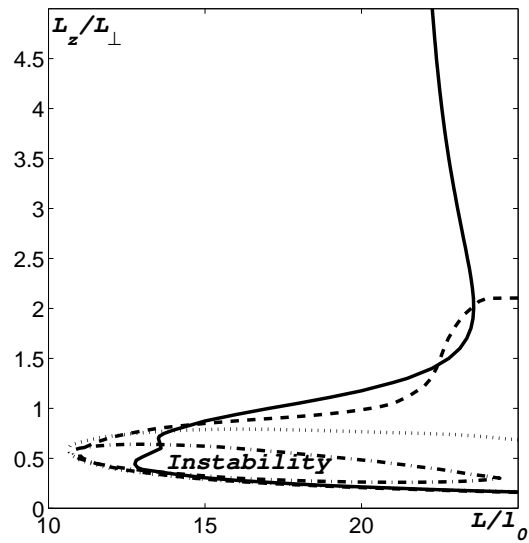}
\caption{\label{FIG6} Range of parameters ($L_z / L_{\perp}$; $L /
l_0$) for which the convective-shear instability occurs, for
different values of the angle  $\varphi$ between the horizontal wave
vector and the $x$-axis: $\varphi = 0^\circ$ (solid line); $\varphi
= 18^\circ$ (dashed line); $\varphi = 30^\circ$ (dotted line);
$\varphi = 90^\circ$ (dashed-dotted line). Here $\alpha=1$ and
$\lambda=0.1$\,.}
\end{figure}

\begin{figure}
\centering
\includegraphics[width=7cm]{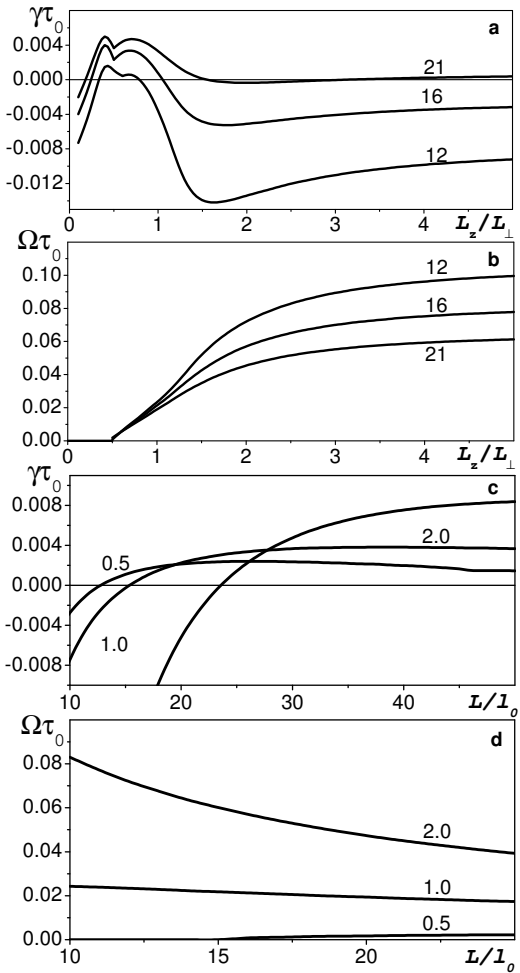}
\caption{\label{FIG7} Growth rates of the convective-shear
instability and the frequencies of the generated convective-shear
waves versus: $L_z / L_{\perp}$ and $L / l_0$. Corresponding
dependencies on the parameters $L / l_0$ are given for different
$L_z / L_{\perp}$ and vice versa. Here $\alpha=1$ and
$\lambda=0.1$\,.}
\end{figure}

Let us consider an evolution of perturbations with zero
$y-$derivatives of the fields $\tilde U, \, \tilde W, \, \tilde S$.
We seek for a solution of Eqs.~(\ref{B3})-(\ref{B5}) in the form $
\propto \exp(\gamma \, t - i \, {\bf K} \cdot {\bf r}) $. The growth
rate of the convective-shear instability of long-wave perturbations
is given by \cite{EKRZ02,EKRZ06}
\begin{eqnarray}
\gamma \propto g \, F_z^\ast \, \tau_0^2 \, K^{2} \, (\beta \,
\lambda \, \sin^2 \theta)^{2/3} \;, \label{B6}
\end{eqnarray}
where $\lambda = \sigma \, \tau_0$ is the shear parameter, the
parameter $\beta \equiv (l_{0} K)^{-2} \gg 1$ and $K = \sqrt{K_{x}^2
+ K_{z}^2} $. The convective-shear instability causes formation of
large-scale coherent fluid motions in the form of rolls (see
Fig.~\ref{FIG4}) aligned along the imposed mean velocity $ {\bf
U}^{(0)} .$ The instability can also result in generation of the
convective-shear waves with the frequency
\begin{eqnarray}
\Omega \propto \sqrt{3} \, g \, F_z^\ast \, \tau_0^2 \, K^{2} \,
(\beta \, \lambda \, \sin^2 \theta)^{2/3} \;, \label{B7}
\end{eqnarray}
which implies the wave-number dependence, $\Omega \propto K^{2/3}$.
The convective-shear waves propagate perpendicular to convective
rolls (see Fig.~\ref{FIG4}). The analysis of the convective-shear
instability was performed in Refs. \cite{EKRZ02,EKRZ06} only for a
small square of Brunt-V\"{a}is\"{a}l\"{a} frequency and zero
$y-$derivatives of the fields $\tilde U, \, \tilde W, \, \tilde S$.
This corresponds to the convective-shear instability for a very
small component of the wave number along the imposed mean shear
(i.e., uniform perturbations along the large-scale shear velocity).
In this case the growth rate of the convective-shear instability is
maximum.

In the present study we consider arbitrary values of the
Brunt-V\"{a}is\"{a}l\"{a} frequency and perform the numerical
analysis of the convective-shear instability for nonzero
$y-$derivatives of the fields $\tilde U, \, \tilde W, \, \tilde S$.
We consider the eigenvalue problem with boundary conditions. We seek
for a solution of Eqs.~(\ref{B3})-(\ref{B5}) in the form $ \propto
\Psi(z) \exp(\gamma \, t - i \, {\bf K}_\perp \cdot {\bf r}) $,
where the eigenfunction $\Psi(z)$ and the growth rate $\gamma$ of
the convective-shear instability are determined by
Eqs.~(\ref{B3})-(\ref{B5}). The system of the ordinary differential
equations for the eigenvalue problem is solved numerically with the
following boundary conditions: $\tilde U = \tilde U'' = \tilde
U^{IV} = \tilde W = \tilde S = 0$ at $z=0$, and $\tilde U' = \tilde
W' = \tilde S' = 0$ at $z=1$, where $f' = df /dz$. We also take into
account that for a turbulent convection, the turbulent Prandtl
number can be estimated as ${\rm Pr}_{_{\rm T}}^{-1} \approx 4 /
(1+{\rm Pr}) \approx 2.34$ with ${\rm Pr}=0.71$ (for air flow). The
latter estimate follows from the balance equations for the turbulent
heat flux, the entropy fluctuations and the turbulent kinetic energy
\cite{EKRZ02}.

\begin{figure}
\centering
\includegraphics[width=7cm]{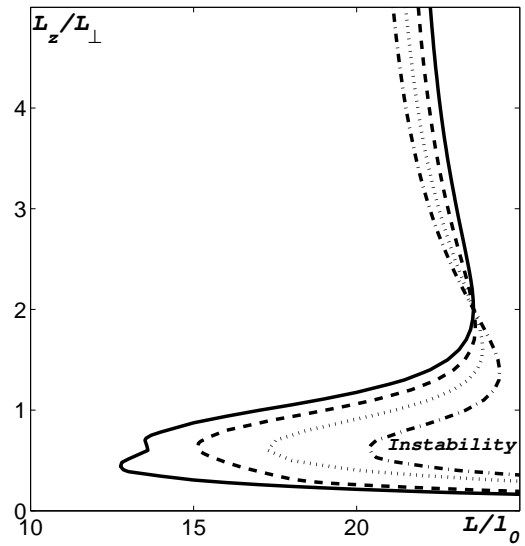}
\caption{\label{FIG8} Range of parameters ($L_z / L_{\perp}$; $L /
l_0$) for which the  convective-shear instability occurs, for
$\lambda=0.1$ for different values of the degree of thermal
anisotropy $\alpha$: $\, \, \, \alpha = 1$ (solid line); $\alpha =
0.9$ (dashed line); $\alpha = 0.8$ (dotted line); $\alpha = 0.7$
(dashed-dotted line).}
\end{figure}

\begin{figure}
\centering
\includegraphics[width=7cm]{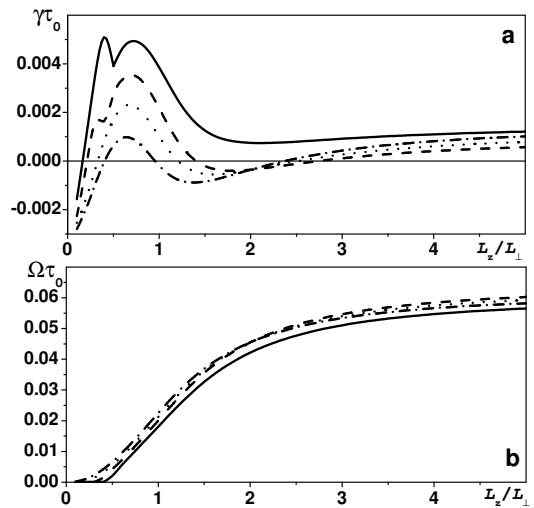}
\caption{\label{FIG9} Growth rates of the convective-shear
instability and the frequencies of the generated convective-shear
waves for different values of the degree of thermal anisotropy
$\alpha$: $\, \, \, \alpha = 1$ (solid line); $\, \, \, \alpha =
0.9$ (dashed line); $\alpha = 0.8$ (dotted line); $\alpha = 0.7$
(dashed-dotted line). Here $\lambda=0.1$ and $L / l_0 = 23$\,.}
\end{figure}

\begin{figure}
\centering
\includegraphics[width=7cm]{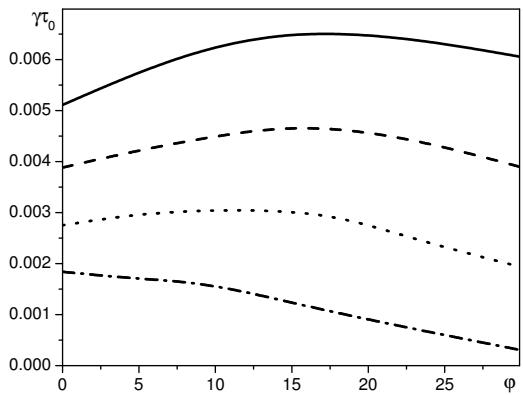}
\caption{\label{FIG10} Growth rates of the convective-shear
instability versus the angle $\varphi$ between the horizontal wave
vector and the $x$-axis for different values of the degree of
thermal anisotropy $\alpha$: $\, \, \, \alpha = 1$ (solid line); $\,
\, \, \alpha = 0.9$ (dashed line); $\alpha = 0.8$ (dotted line);
$\alpha = 0.7$ (dashed-dotted line). Here $\lambda=0.1$, the values
$L_z / L_{\perp}$ and $L / l_0$ correspond to maximum growth rates
of the instability.}
\end{figure}

\begin{figure}
\centering
\includegraphics[width=7cm]{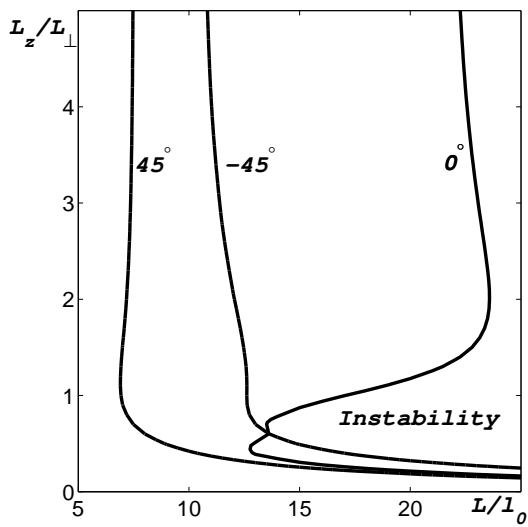}
\caption{\label{FIG11} Range of parameters ($L_z / L_{\perp}$; $L /
l_0$) for which the convective-shear instability occurs, for
different directions $\psi$ of the lateral background heat flux: $\,
\, \, \psi = -45^\circ$; $\psi = 0^\circ$; $\psi = 45^\circ$. Here
$\lambda=0.1$\,.}
\end{figure}

\begin{figure}
\centering
\includegraphics[width=7cm]{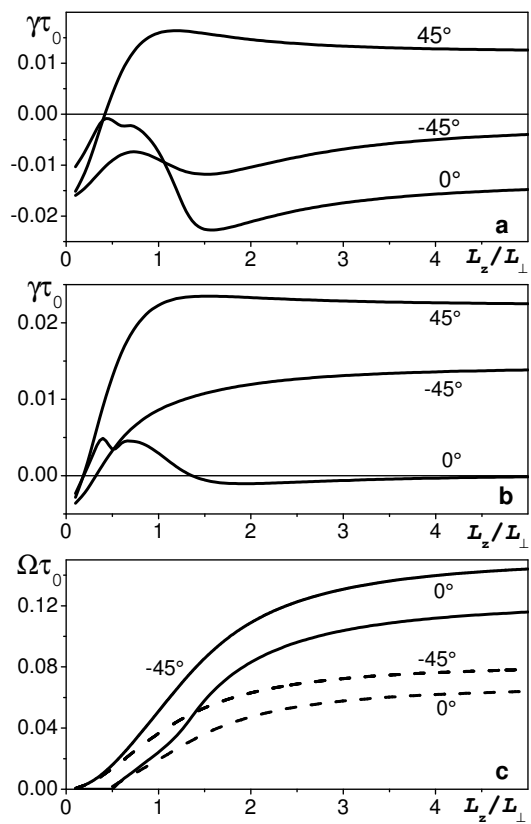}
\caption{\label{FIG12} Growth rates of the the convective-shear
instability and the frequencies of the generated convective-shear
waves versus $L_z / L_{\perp}$ for different directions $\psi$ of
the lateral background heat flux: $\, \, \, \psi = -45^\circ$; $\psi
= 0^\circ$; $\psi = 45^\circ$. (a). The growth rates of the
instability for $L/l_0=10$; (b). The growth rates of the instability
for $L/l_0=20$; (c). The frequencies of the generated waves for
$L/l_0=10$ (solid line) and $L/l_0=20$ (dashed line). Here
$\lambda=0.1$\,.}
\end{figure}

\begin{figure}
\centering
\includegraphics[width=7cm]{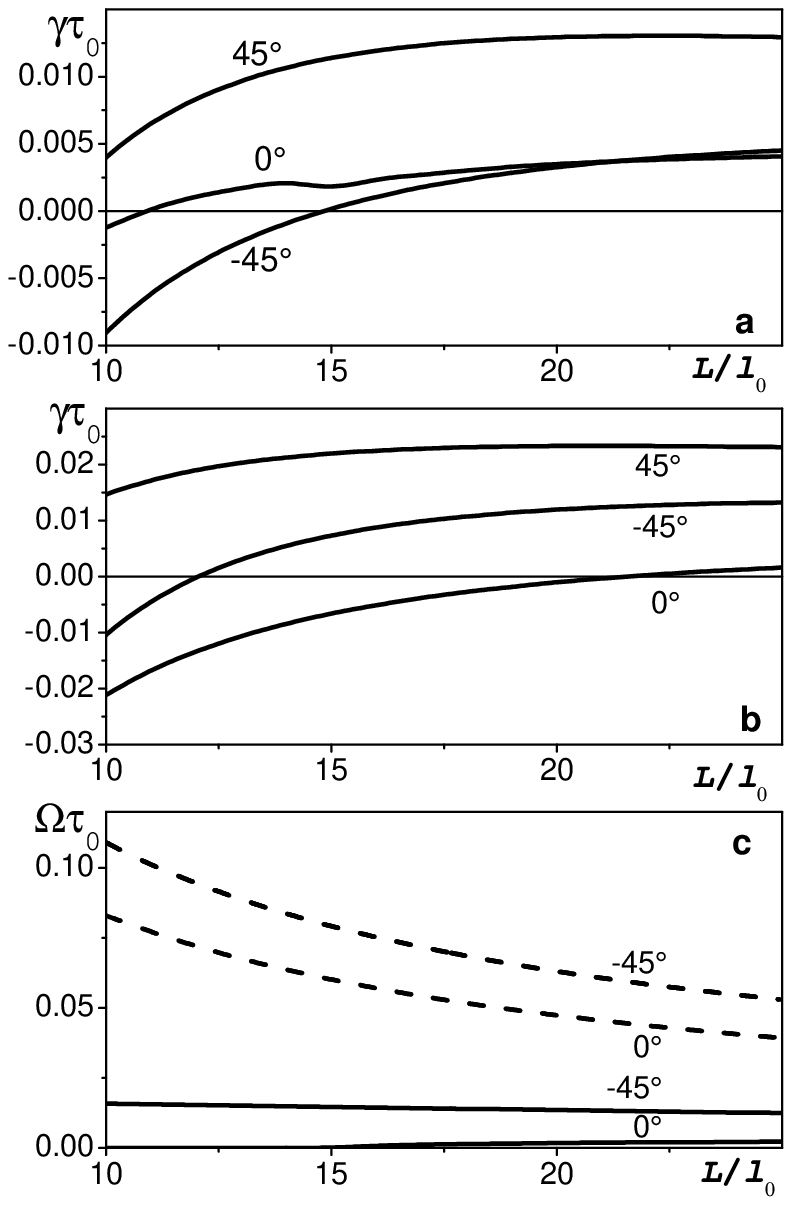}
\caption{\label{FIG13} Growth rates of the the convective-shear
instability and the frequencies of the generated convective-shear
waves versus $L/l_0$ for different directions $\psi$ of the lateral
background heat flux: $\, \, \, \psi = -45^\circ$; $\psi = 0^\circ$;
$\psi = 45^\circ$. (a). The growth rates of the instability for $L_z
/ L_{\perp} = 0.5$; (b). The growth rates of the instability for
$L_z / L_{\perp} = 2$; (c). The frequencies of the generated waves
for $L_z/L_{\perp} = 0.5$ (solid line) and $L_z/L_{\perp} = 2$
(dashed line). Here $\lambda=0.1$\,.}
\end{figure}

Let us consider the thermally isotropic $(\alpha = 1)$ turbulent
convection. Figure~\ref{FIG5} shows the range of parameters ($L_z /
L_{\perp}$; $L / l_0$) for which the convective-shear instability
occurs for different values of the shear parameter $\lambda$. Here
$L\equiv1/\sqrt{L_z^{-2} +L_{\perp}^{-2}}$ and we assumed that
$a_\ast=1$. The case $\lambda=0$ in Fig.~\ref{FIG5} corresponds to
the convective-wind instability (the shear-free turbulent
convection). Inspection of Fig.~\ref{FIG5} shows that the increase
of shear is favorable for the excitation of the convective-shear
instability. In Fig.~\ref{FIG6} we plotted the range of parameters
($L_z / L_{\perp}$; $L / l_0$) for the convective-shear instability
for different values of the angle $\varphi$ between the horizontal
wave vector ${\bf K}_{\perp}$ and the $x$-axis. Increasing the angle
$\varphi$ prevents from the excitation of the convective-shear
instability (i.e., reduces the range of parameters for which the
instability occurs). In Fig.~\ref{FIG7} we plotted the growth rates
of the convective-shear instability and the frequencies of the
generated convective-shear waves versus $L_z / L_{\perp}$ and $L /
l_0$. The curves in Fig.~\ref{FIG7} have a point $L_\ast$ whereby
the first derivative $d\gamma /d K$ has a singularity which is
indicative of bifurcation. The growth rate of the convective-shear
instability for very small $y-$derivatives of the fields $\tilde U,
\, \tilde W, \, \tilde S$, is determined by cubic algebraic equation
\cite{EKRZ02}. Below the bifurcation point, the cubic equation has
three real roots (which corresponds to aperiodic instability without
generation of waves). Above the bifurcation point, the cubic
equation has one real and two complex conjugate roots. In this case
the convective-shear waves are generated. The source of energy for
these waves is the turbulence energy.

Now we perform the detailed numerical analysis of the
convective-shear instability in order to determine the key
parameters that affect this instability. First, we study the effect
of the thermal anisotropy $\alpha$ on the convective-shear
instability. Figure~\ref{FIG8} shows the range of parameters ($L_z /
L_{\perp}$; $L / l_0$) for which the convective-shear instability
occurs, for different values of the thermal anisotropy $\alpha$. In
Fig.~\ref{FIG9} we plotted the growth rates of the convective-shear
instability and the frequencies of the generated convective-shear
waves for different values of $\alpha$. The decrease of the degree
of thermal anisotropy $\alpha$ increases the threshold in the
parameter $L / l_0$ required for the excitation of the
convective-shear instability. Figure~\ref{FIG10} shows the growth
rates of the convective-shear instability versus the angle $\varphi$
between the horizontal wave vector ${\bf K}_{\perp}$ and the
$x$-axis for different values of $\alpha$. Here the values $L_z /
L_{\perp}$ and $L / l_0$ correspond to the maximum growth rates of
the instability. For $\alpha > 0.7$ the growth rate of the
convective-shear instability attains the maximum for $\varphi_m >
0^{\circ}$. An increase of the degree of thermal anisotropy $\alpha$
increases the angle $\varphi_m$. In the thermally isotropic $(\alpha
= 1)$ turbulent convection the angle $\varphi_m = 18^\circ$, while
for $\alpha = 0.8$ (i.e., $\xi \approx 0.92$), the angle $\varphi_m
\approx 10^\circ$. Note that according to the atmospheric
observations, the observed angle between the cloud streets and
direction of the wind is of the order of $10^\circ - 14^\circ$. The
calculated angle $\varphi_m$ is in compliance with these
observations. Note that the convective rolls are stretched in the
horizontal plane in the direction perpendicular to ${\bf K}_\perp$
and the shear velocity is directed along the $y$-axis. Inspection of
Figs.~\ref{FIG9} and \ref{FIG10} shows that decrease of the
parameter $\alpha$ reduces the growth rates of the convective-shear
instability. In Figs.~\ref{FIG8}-\ref{FIG10} we considered the case
$\alpha \leq 1$ which is of interest in view of the atmospheric
applications.

Next, we study the effect of the lateral background heat flux
(determined by the third term $\propto (({\bf F}^{\ast} {\bf \times}
{\bf e}) \cdot \bec{\nabla}) \, \tilde W$ in the right hand side of
Eq.~(\ref{B1})), on the convective-shear instability. We introduce
the angle $\psi$ between the horizontal component ${\bf
F}^{\ast}_{\perp}$ of the background turbulent heat flux and $x$
axes, where the total background heat flux is ${\bf F}^{\ast} =
(F_{\perp}^{\ast} \, \cos \psi, F_{\perp}^{\ast} \, \sin \psi,
F_{z}^{\ast})$. The angle $\psi$ is determined by the boundary
conditions in the horizontal plane (e.g., by the temperature
gradient in the horizontal plane). Figure~\ref{FIG11} shows the
range of parameters ($L_z / L_{\perp}$; $L / l_0$) for which the
convective-shear instability occurs, for different directions $\psi$
of the lateral background heat flux ${\bf F}^{\ast}_{\perp}$. In
Figs.~\ref{FIG12}-\ref{FIG13} we plotted the growth rates of the
convective-shear instability and the frequencies of the generated
convective-shear waves for this case, where $F_{\perp}^{\ast} /
F_{z}^{\ast} = 0.5$. Note that the background mean vorticity due to
the imposed large-scale shear is ${\bf W}^{(0)} = \bec{\nabla} {\bf
\times} {\bf U}^{(0)} = - \sigma \, {\bf e}_x$. This is the reason
why there is no symmetry with respect to the $YZ$ plane of the
large-scale shear, i.e., the contributions to the convective-shear
instability caused by the positive and negative angles $\psi$ of the
lateral background heat flux are different. In particular, the range
of the convective-shear instability in the presence the lateral
background heat flux with the positive angles $\psi$ is wider than
that for the negative angles $\psi$ (see Fig.~\ref{FIG11}). On the
other hand, even for the negative angles $\psi$ the range of the
convective-shear instability is wider than that in the absence of
the lateral background heat flux. Note also that in the presence of
the lateral background heat flux with the positive angles $\psi$,
the convective-shear waves are not generated. This is reason why we
plotted in Figs.~\ref{FIG12}c-\ref{FIG13}c the frequencies of the
generated convective-shear waves only for $\psi \leq 0$.

Note that there are three groups of parameters in this study of the
large-scale coherent structures formed in a turbulent convection:

(i) the external parameters: the value of shear $\sigma$ and the
background heat flux ${\bf F}^{\ast} = (F_{\perp}^{\ast} \, \cos
\psi, F_{\perp}^{\ast} \, \sin \psi, F_{z}^{\ast})$;

(ii) the parameters which determine the background turbulent
convection: the degree of thermal anisotropy $\alpha$, the
correlation time $\tau_0 = l_0 / u_0$ and the parameter $a_{\ast}=2
\, g \, \tau_0 \, F_z^{\ast} / u_0^2$;

(iii) the parameters related to the characteristics of the
large-scale coherent structures: the aspect ratio of the structure
$L_{\perp}/ L_z$, the minimum size  $L$ of the structure, $L_x =
L_{\perp} \, \cos \varphi$, the characteristic time of the formation
of the large-scale coherent structures $\propto \gamma^{-1}$, and
the frequency $\Omega$ of the generated convective-shear waves.

The parameters related to the characteristics of the large-scale
coherent structures are determined in this study. The external
parameters and the parameters which determine the background
turbulent convection at the present level of analysis are treated as
free parameters. The external parameters are determined by the
boundary conditions. The degree of thermal anisotropy $\alpha$ can
be determined by the budget equation for the two-point correlation
function for the velocity-entropy fluctuations. This parameter has
been recently measured in a laboratory experiment in turbulent
convection in air-flow \cite{EEKR06}. In the range of the Rayleigh
numbers $10^7 - 10^8$ (based on the kinematic viscosity and
molecular diffusivity) this parameter varies within the range from
$0.5$ to $2$. The parameter $a_{\ast}$ and the correlation time
$\tau_0 = l_0 / u_0$ can be determined from the budget equations for
the turbulent kinetic energy and vertical turbulent fluxes of
momentum and the entropy. The turbulent correlation time $\tau_0$
and correlation length $l_0$ are measured in laboratory convection
(see, e.g., Ref. \cite{EEKR06}).

\section{DISCUSSION}

In the present study we investigated formation of large-scale
coherent structures in a  non-rotating turbulent convection due to
an excitation of large-scale instabilities. In the shear-free
turbulent convection, the cell-like structures are formed due to the
convective-wind instability. The redistribution of the turbulent
heat flux due to the non-uniform large-scale motions causes strong
reduction of the critical effective Rayleigh number required for the
excitation of the convective-wind instability. The effective
Rayleigh number is based on the eddy viscosity and turbulent thermal
conductivity. We also found that the critical effective Rayleigh
number increases with the increase of the anisotropy of turbulent
temperature diffusivity caused by the buoyancy effects.

In the sheared turbulent convection, the roll-like structures
stretched along the imposed large-scale sheared velocity are formed
due to the large-scale convective-shear instability. This
instability produces the convective-shear waves propagating
perpendicular to the convective rolls. We studied numerically the
convective-shear instability and determined the key parameters that
affect the formation of the large-scale coherent structures in the
turbulent convection. In particular, we found that the degree of
thermal anisotropy and the lateral background heat flux strongly
modify the growth rates of the large-scale convective-shear
instability, the frequencies of the generated convective-shear waves
and change the instability thresholds.

The results described in this study are based on the linearized
mean-field equations, and therefore, they cannot describe detail
features of the turbulent convection observed in the numerous
laboratory experiments
\cite{KH81,SWL89,CCL96,K01,NSS01,NS03,BKT03,XL04,SQ04,BNA05} and in
direct numerical simulations \cite{HTB03,PHP04}. In particular, we
made the following assumptions about the turbulent convection. We
considered a homogeneous, incompressible background turbulent
convection (i.e., the turbulent convection without mean-velocity
gradients). The nonuniform mean velocity affects the background
turbulent convection, i.e., it causes generation of the additional
strongly anisotropic velocity fluctuations by tangling of the
mean-velocity gradients with the background turbulent convection. We
assumed that the generated anisotropic fluctuations do not affect
the background turbulent convection. This implies that we considered
a one-way coupling due to a weak inhomogeneity of the large-scale
velocity. Thus, we studied simple physical mechanisms to describe an
initial stage of the formation of large-scale coherent structures in
a non-rotating turbulent convection. The simple model considered in
our paper can only mimic the real flows associated with laboratory
turbulent convection. Clearly, the comprehensive theoretical and
numerical studies are required for quantitative description of the
laboratory turbulent convection.

In spite of this very simple model it reproduces some properties of
the semi-organized structures observed in the atmospheric turbulent
flows \cite{EKRZ06}. The semi-organized structures are observed in
the form of rolls (cloud streets) or three-dimensional convective
cells (cloud cells). The observed angle between the cloud streets
and the mean horizontal wind of the sheared turbulent convection is
about $10^{\circ}-14^{\circ}$, the lengths of the cloud streets vary
from $20$ to $200$ km, the widths from $2$ to $10$ km, and
convective depths from $2$ to $3$ km. The ratio of the minimal size
of the structure to the maximum scale of turbulent motions $L /
l_{0} = 10 - 100$. The characteristic life time of rolls varies from
$1$ to $72$ hours. Rolls may occur over water surface or land
surfaces \cite{EB93,AZ96}. Our study yield the following parameters
of the convective rolls: $L / l_{0} = 10 - 100$, the characteristic
time of formation of the rolls $\sim \tau_0 / \gamma$ varies from
$1$ to $3$ hours. The life time of the convective rolls is
determined by a nonlinear evolution of the convective-shear
instability. The latter is a subject of a separate ongoing study. We
have shown that the maximum growth rate of the convective-shear
instability is attained when the angle between the cloud streets and
the mean horizontal wind of the convective layer is about
$10^{\circ}-17^{\circ}$ in agreement with observations. We also
found an excitation of the convective-shear waves propagating
perpendicular to convective rolls. This finding is in agreement with
observations in the atmospheric convective boundary layer, whereby
the waves propagating perpendicular to cloud streets have been
detected \cite{BR99}. In addition, the motions in the convective
rolls have a nonzero helicity in agreement with predictions made in
Ref. \cite{ET85}.

There are two types of cloud cells in the atmospheric shear-free
turbulent convection: open and closed. Open-cell circulation has
downward motion and clear sky in the cell center, surrounded by
cloud associated with upward motion. Closed cells have the opposite
circulation \cite{AZ96}. Both types of cells have diameters ranging
from $10$ to $40$ km, they occur in a convective layer with a depth
of about $1$ to $3$ km and the characteristic life time of cloud
cells is about several hours. Our analysis shows that the minimum
threshold value of the effective Rayleigh number required for the
excitation of the large-scale instability is attained at $L_\perp /
L_z = 2$ (see Figs. 1-3, dotted and dashed-dotted curves), is in
agreement with numerous observations. The ratio of the minimum size
of the structure to maximum scale of turbulent motions $ L / l_{0} =
5 - 15$.  The characteristic time of formation of the convective
cells $ \sim \tau_0 / \gamma$ varies from $1$ to $3$ hours.
Therefore, the predictions of the developed theory are in an
agreement with observations of the semi-organized structures in the
atmospheric convective boundary layer. The typical temporal and
spatial scales of structures are always much larger then the
turbulence scales. This justifies separation of scales which was
assumed in the suggested in the theory. Note that the applicability
of the mean-field equations for study of turbulent convection was
discussed in Ref. \cite{TB94}.

In our study we consider non-rotating turbulent convection and apply
our results to the atmospheric convective boundary layers, where the
shear is usually caused by wind. The rotation of the Earth usually
affects the hight of the atmospheric convective boundary layer. The
rotation can also affect the longitudinal spatial structure of the
cloud streets. Note that in astrophysical applications the shear (or
differential rotation) can in general be a consequence of
anisotropies in rotating systems. Our study can be also useful for
understanding the origin of formation of the meso-granular
structures in the solar convection (see \cite{CLW01}).

\begin{acknowledgments}
The authors benefited from stimulating discussions with F.H.~Busse,
D.~Etling, H.J.S.~Fernando, R.~Foster, A.~Tsinober and
S.~Zilitinkevich. This work was partially supported by the Israel
Science Foundation governed by the Israeli Academy of Science and
the Israeli Universities Budget Planning Committee (VATAT).

\end{acknowledgments}

\end{document}